\documentclass[12pt]{article}
\usepackage{graphicx}
\begin{document}
\centerline {\large \bf Frustration from Simultaneous Updating in Sznajd 
Consensus Model}

\bigskip
\centerline{Dietrich Stauffer}

\centerline{Institute for Theoretical Physics, Cologne University}

\centerline{D-50923 K\"oln, Euroland}

\bigskip
{\small
In the Sznajd model of 2000, a pair of neighbouring agents on a square lattice
convinces its six neighbours of the pair opinion iff the two agents of the pair 
share the same opinion. Now we replace the usual random sequential updating
rule by simultaneous updating and find that this change makes a complete
consenus much more difficult. The larger the lattice is, the higher must be the
initial majority for one of the two competing opinions to become the consensus.}
\bigskip


\bigskip
The application of cellular automata, Ising models and other tools of 
(computational or statistical) physics has a long tradition (Majorana
1942, Schelling 1971, Sakoda 1971, Callen and Shapero 1974, Galam et al 1981,
Schweitzer 1997, Weidlich 2000). In particular one would like to know the
conditions to reach a consensus out of an initially diverging set of 
opinions (Deffuant et al 2000, Kobayashi 2001, Hegselmann and Krause 2002).
Most models assume that every agent is influenced by its neighbours and takes,
for example, the opinion of the majority of them, or of a weighted average.
The Sznajd model (Sznajd-Weron and Sznajd 2000; for a review see Stauffer 2002),
on the other hand, assumes that every agent tries to influence its neighbours,
without caring much about what they think first. Thus in the Sznajd model
the information flows outward to the neighbourhood, as in infection or rumour 
spreading (Noymer 2001), while in most other models the information flows 
inward from the neighbourhood. Also, the Sznajd model takes into account the
well-known psychological and political fact that ``united we stand, divided we
fall''; only groups of people having the same opinion, not divided groups, can
influence their neighbours. 

On the square lattice, where every site is occupied by an agent having one
of two possible opinions +1 and $-$1, the most-studied Sznajd rule is:
{\bf A pair of nearest neighbours convinces its six nearest neighbours of the
pair opinion if and only if both members of the pair have the same opinion; 
otherwise the pair and its neighbours do not change opinion.} Initially
the opinions are distributed randomly, +1 with probability $p$ and $-1$ with 
probability $1-p$. This standard model then gave always a consensus, which for
large lattices was that opinion which initially had a majority; if $p = 1/2$
initially, then half of the cases ended with everybody having opinion +1, and 
the other half of the cases with the opposite opinion.

In these simulations random sequential updating was used, i.e. one of the
$L \times L$ agents in the square lattice was selected randomly, and then one 
of its four neighbours to check if they share the same opinion. One time step
was completed if on average each of the $L \times L$ agents was selected once
as the first member of the pair. A. Iosselevitch (private communication)
suggested to compare this rule with the simultaneous updating traditional for
cellular automata, used also in the consensus model of Kobayashi (2001): Each
pair is judged by its opinion at time $t$ to gives its six neighbours their
possibly new opinions at time $t+1$. Now we can go through the lattice like a
typewriter to find the first member of the pair; only for the second member
of the pair a random selection is still needed. Going through the whole 
lattice once constitutes one time step. 

In both random sequential as well as simultaneous updating, an agent can 
belong to the neighbourhoods of several convincing pairs. For random sequential
updating, the agent 
then follows each pair in the order in which it receives orders,
just like civil servants followed their various governments in Germany during 
the 20th century. For simultaneous updating, on the other hand, it does not
know what to do if one pair has opinion +1 and another also neigbouring pair 
has the opposite opinion. It then feels frustrated and does nothing. i.e. it
stays with its old opinion. (Similar frustration effects are known from some
models of magnetism.) This frustration then hinders the development of a
consensus.

With up to 800 samples, and $L \ge 13$ we never found a consensus at $p = 1/2$.
One has to use small lattices, or $p$ different from 1/2, to find all agents
at the end having the same opinion. Fig.1 shows how the number (among 800) of
samples without a consensus even after 10,000 time steps varies with lattice
size $L$ and initial concentration $p$; if there was a consensus it was in 
favour of the initial majority. (For $7 \le L < 13$ also at $p=1/2$ rare cases 
of a consensus were found.) Fig.2 shows how the $p$ needed to get a consensus 
in half the cases varied for varying lattice size; the problem is by definition
symmetric about $p = 1/2$, and only $p < 1/2$ is thus plotted in our figures.

Of course, reality differs from a square lattice. If everybody interacts with 
everybody else equally, without regard of a geometrical distance (Kobayashi
2001), this ``mean field''  problem may be suited for analytical solution or
description by differential equation. More realistic may be a network where
a few agents have lots of neighbours, and most have only a few neighbours, 
without a sharp boundary between celebrities and common folks (Albert and
Barabasi 2002). Here an Ising model gave already an unusual phase transition
(Aleksiejuk et al 2002), and a different Sznajd model with many possible 
opinions agreed with Brazilian election results.

Our simultaneous updating corresponds to formal committee meetings at times
fixed for all participants, while random sequential updating corresponds to
informal meetings of subgroups at various times. Our simulations then indicate 
that informal meetings have a higher chance to lead to a consensus. 
 
\begin{figure}[hbt]
\begin{center}
\includegraphics[angle=-90,scale=0.5]{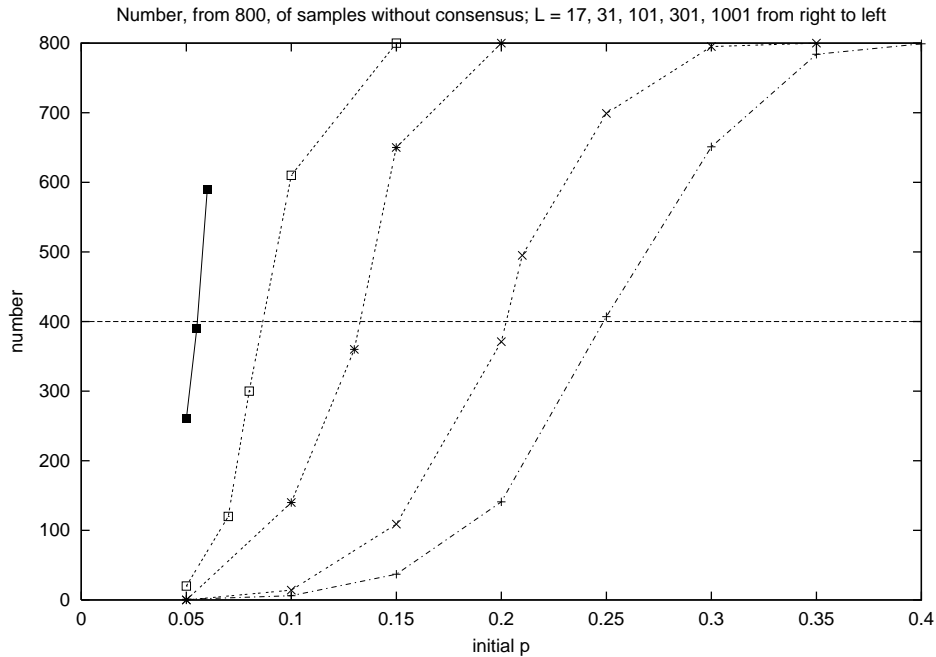}
\end{center}
\caption{Number, from 800 samples, of cases where still a consensus was 
reached, for $L = 17$ and 31. For $L = 101$, 301, and 1001 only 80 samples
were run and the resulting numbers thus multiplied by 10.
}
\end{figure}

\begin{figure}[hbt]
\begin{center}
\includegraphics[angle=-90,scale=0.5]{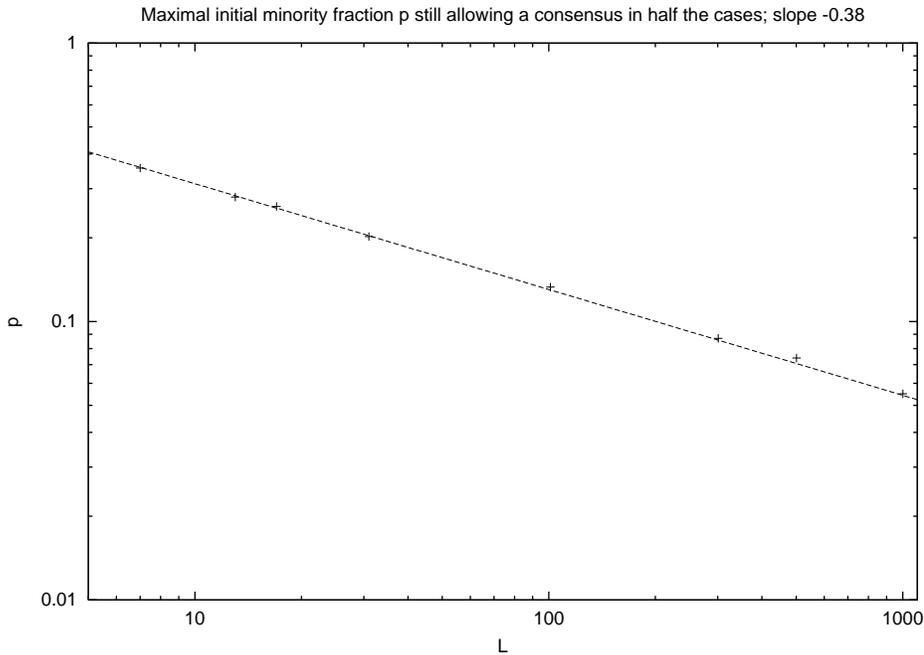}
\end{center}
\caption{Variation with $L$ of the initial probability for which in half the 
cases a consensus was reached. 
}
\end{figure}

I thank A. Iosselevitch for suggesting this work and the supercomputer center in
J\"ulich for time on their Cray-T3E.

\bigskip
\parindent 0pt
R. Albert, A.L. Barabasi, Statistical mechanics of complex networks.
Reviews of Modern Physics 74: 47-97 (2002).

A. Aleksiejuk, J.A. Ho{\l}yst and D. Stauffer.
Ferromagnetic Phase Transition in Barab\'asi-Albert Networks. Physica A 310: 
260-266 (2002)

E. Callen and D. Shapero, A Theory of Social 
Imitation, Physics Today, July 1974, 23-28.

G. Deffuant, D. Neau, F. Amblard and G. Weisbuch, Mixing beliefs among 
interacting agents, Adv. Complex Syst. 3: 87-98 (2000).

S. Galam, Y. Gefen and Y. Shapir, Sociophysics: A mean behavior model for the 
process of strike, J. Mathematical Sociology 9: 1-13 (1982).

R. Hegselmann and M. Krause, Opinion dynamics and bounded confidence,
for Journal of Artificial Societies and Social Simulation 5 (2002).

J. Kobayashi, Unanimous Opinions in Social Influence Networks. J. Mathematical 
Sociology 24: 285-297 (2001).

E. Majorana, Il valore delle leggi statistiche nella fisica e nelle scienze 
sociali, Scientia {\bf 36}: 58-66 (1942).

A. Noymer, The transmission and persistence of `urban legends': Sociological
application of age-structured epidemic models. J. Mathematical Sociology 25:
299-323 (2001).

F. Schweitzer, (ed.) {\it  Self-Organization of Complex Structures:
From Individual to Collective Dynamics}, Gordon and Breach, Amsterdam 1997.

J.M. Sakoda, The checkerboard model of social interaction,
J. Mathematical Sociology 1: 119-132 (1971).

T.C. Schelling, Dynamic Models of Segregation, J. Mathematical Sociology 1:
143-186 (1971). 

D. Stauffer, Monte Carlo Simulations of the Sznajd model, Journal of 
Artificial Societies and Social Simulation 5, No.1, paper 4 (2002) 
(jasss.soc.surrey.ac.uk).

K. Sznajd-Weron and J. Sznajd, Opinion Evolution in Closed Community.
Int. J. Mod. Phys. C 11: 1157-1166 (2000). 

W. Weidlich, {\it Sociodynamics; A Systematic Approach to Mathematical Modelling
in the Social Sciences}. Harwood Academic Publishers, 2000.
\end{document}